\documentclass[APA,STIX1COL]{WileyNJD-v2}
\usepackage[section]{placeins}
\usepackage{natbib}
\articletype{}%

\begin{document}

\title{Vector Time Series Modelling of Turbidity in Dublin Bay}

\author[1]{Amin Shoari Nejad*}

% \author[2]{Sarah E. Heaps}

\author[2]{Gerard D. McCarthy}

\author[3]{Brian Kelleher}

\author[3]{Anthony Grey}

\author[1]{Andrew Parnell}

\authormark{SHOARI NEJAD \textsc{et al}}

\address[1]{Hamilton Institute, Insight Centre for Data Analytics, Maynooth University, Kildare, Ireland}

\address[2]{
ICARUS, Department of Geography, Maynooth University, Maynooth, Ireland}

\address[3]{
Organic Geochemical Research Laboratory, Dublin City University, DCU Glasnevin Campus, Dublin 9, Ireland}

\corres{*Amin Shoari Nejad, \email{amin.shoarinejad@gmail.com}}

%\presentaddress{This is sample for present address text this is %sample for present address text}

\abstract[Abstract]{Turbidity is commonly monitored as an important water quality index. Human activities, such as dredging and dumping operations, can disrupt turbidity levels and should be monitored and analyzed for possible effects. In this paper, we model the variations of turbidity in Dublin Bay over space and time to investigate the effects of dumping and dredging while controlling for the effect of wind speed as a common atmospheric effect. We develop a novel Vector Auto-Regressive Conditional Heteroskedasticity (VARCH) approach to modelling the dynamical behaviour of turbidity over different locations and at different water depths. We use daily values of turbidity during the years 2017-2018 to fit the model. We show that the results of our fitted model are in line with the observed data and that the uncertainties, measured through Bayesian credible intervals, are well calibrated. Furthermore, we show that the daily effects of dredging and dumping on turbidity are negligible in comparison to that of wind speed.}

\keywords{Bayesian, Vector Autoregression, Turbidity}

%\jnlcitation{\cname{%
%\author{Williams K.}, 
%\author{B. Hoskins}, 
%\author{R. Lee}, 
%\author{G. Masato}, and 
%\author{T. Woollings}} (\cyear{2016}), 
%\ctitle{A regime analysis of Atlantic winter jet variability applied to %evaluate HadGEM3-GC2}, \cjournal{Q.J.R. Meteorol. Soc.}, %\cvol{2017;00:1--6}.}

\maketitle

%\footnotetext{\textbf{Abbreviations:} ANA, anti-nuclear antibodies; APC, antigen-presenting cells; IRF, interferon regulatory factor}

\section{Introduction}\label{sec1}

Studying the variables affecting turbidity is of importance in maintaining coastal ecosystem health. Turbidity is an index for water clarity which measures how suspended solids in water hinder the transmission of light \citep{davies2001}. There are many sources of suspended solids including: phytoplankton; particles from coastal erosion; re-suspended bed sediments; organic detritus from streams; and excessive algae growth \citep{Briciu2014}. Variability in water turbidity influences the transportation dynamics and distribution of nutrients, contaminants, and biological production \citep{TIAN2009, kennish1991, diez2020, ganju2020, GE2020}. Water turbidity is an important habitat factor in many estuarine systems, and changes in it can have a significant impact on management decisions such as the dredging of ports and canals \citep{bever2018}. 

Our goal in this paper is to evaluate the variations of turbidity in Dublin Bay explained by dredging and dumping operations when controlling for the effect of wind speed, which is an important atmospheric contributor. Dublin has a long history of difficult access for ships to the port area due to sandbanks at the mouth of the port \citep{DPC2016}. To solve this problem regular dredging operations have been carried over decades to remove unwanted waste as well as dangerous accumulations of sediments from areas that ships use when entering the port. The excavated materials from the dredging operations are dumped at a more remote location in the bay.   

There are relatively few studies focusing on water turbidity in Dublin bay. In one example, \cite{Briciu2014} used frequentist statistical tests to show that turbidity can be strongly influenced by vessel activity in Dublin bay using data collected from a single location. By contrast, we take a broader approach and look at multiple measuring sites simultaneously corresponding to both the sites where sediment is dumped and dredged, whilst considering issues of turbidity down the water column. We develop a novel Vector Auto Regressive Conditional Heteroskedasticity (VARCH) model to control for the spatio-temporal structure using turbidity data measured by five buoys installed at different locations in the bay. Our model combines three well known approaches:  Bayesian time series modelling, which has been successfully applied to many problems \citep[for example see][]{chao2020, hensman2013,  senf2017}; vector autoregression (VAR), originally introduced by \cite{sims1980} and widely used in macroeconomics, causal inference, and forecasting \citep{karlsson2013, alola2021, santos2022bayesian}; and autoregressive conditional heteroscedasticity (ARCH) models that have been employed in many forecasting applications \citep[see e.g.][]{degiannakis2004}. In particular, Bayesian time series modelling with a VAR structure has been used before to model spatio-temporal data. For example, \cite{santos2022bayesian} used a Bayesian VAR approach to model stream networks time series that were collected from different locations. In this paper, we extend the Bayesian VAR models by relaxing the assumption of constant variance over time, and letting the data generating process incorporate time and space varying variance. To the best of our knowledge, there have been no attempt in building this type of model for real world data which are most likely heteroscedastic in nature.

We organise our paper as follows. In Section \ref{data}, we describe the data we use in our study. In Section \ref{modelling}, we explain our modelling framework. In Section \ref{sec4}, we discuss our findings including plots of the model outputs. We summarise the paper in Section \ref{disc} by considering the strengths and weaknesses of our approach and potential areas for future research.

\section{Data description}\label{data}

\begin{figure}[htbp]
\centerline{\includegraphics[scale=.4]{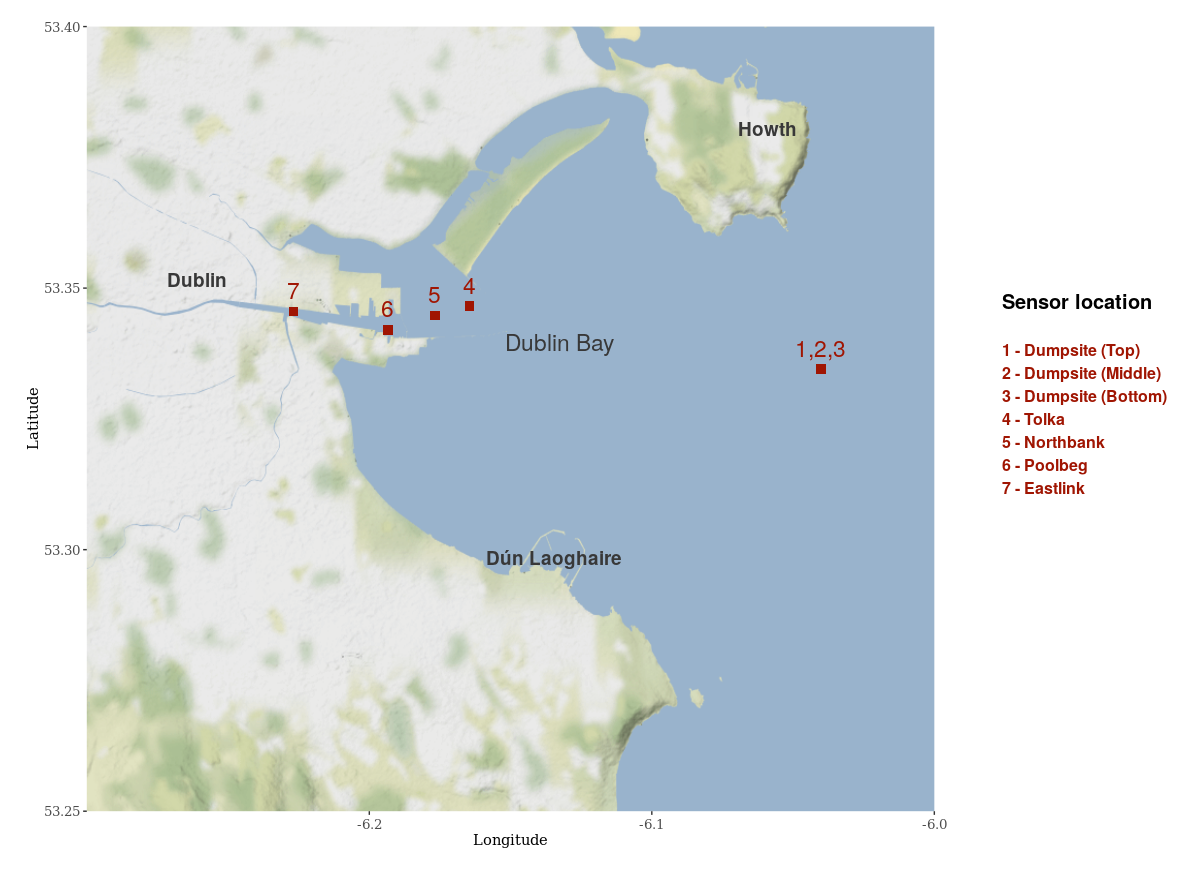}}
\caption{Buoys measuring turbidity in Dublin Bay. We use the same numbering scheme when referring to each site throughout the paper. Buoys 4 to 7 are potential dredging sites, whilst the sediment is dumped at dumpsite.}
\label{fig1}
\end{figure}

Water turbidity levels are measured in Nephelometric Turbidity Units (NTU) which calculate the amount of light reflected through a set of suspended particles. Our dataset contains measurements of water turbidity in NTU at five different locations, four of which take measurements at a single depth and are located throughout the channel from the River Liffey towards Dublin Bay where dredging takes place. The fifth buoy takes measurements at three different levels of the water column and is located approximately 10 kilometres away from Dublin port at the location where the sediments are dumped.  Figure \ref{fig1} shows the locations of the buoys in the bay. 

\begin{figure}[htbp]
\centerline{\includegraphics[scale=.4]{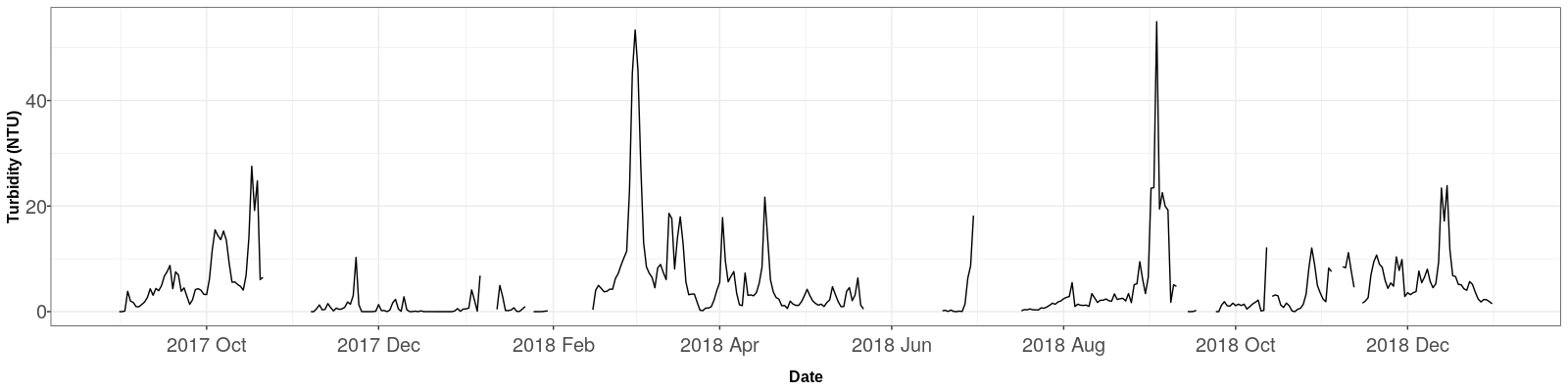}}
\caption{Daily measurements of turbidity at Tolka from  31/08/2017 to 31/12/2018.}
\label{raw_plt}
\end{figure}

Turbidity measurements are recorded every 15 minutes by the buoys, but for our analysis we aggregated the raw data into daily averages. This allowed us to focus on the impact of dredging whilst removed short term fluctuations (e.g. that of tides) or the instantaneous impact of vessels arriving or leaving from the port. The aggregation resulted in a total of 488 daily observations per buoy from 31/08/2017 to 31/12/2018. However there are some periods with missing data which seems to be due to equipment failure (e.g. discharged batteries) and gives rise to data gaps when working with our sensor data. A plot of the raw data with missing values at one buoy is provided in  Figure \ref{raw_plt}. Additionally, we use wind speed data measured at Dublin Airport, provided by \cite{winddata} for the same period as turbidity data, to control for its effect on turbidity. 

\section{Modelling procedure}\label{modelling}

In this section we describe the general modelling framework that we follow to build a dynamic spatio-temporal model that describes the response of turbidity to a variety of environmental factors. We then provide specific variations on this template to create four different models which we use for fitting on the data.  We denote $Y_{t}$ as an $S$-vector of turbidity measurements at time $t$ where $S$ is the number of locations (or equivalently buoys), $s = 1,2, ..., S$ represent the locations and times $t \ (t= 1, 2, ..., T)$. We write the model hierarchically in two main layers as: 

\begin{eqnarray}
     Y_{t} \mid M_{t}, \Sigma_t  &\sim& MVN\left(M_{t}, \Sigma_{t} \right)  \\
     M_{t} &=& A + \sum_{j=1}^{P} X_{jt} \circ \beta_{j} + U_{t}
\end{eqnarray} 

where $M_t$ is the process mean and $\Sigma_{t}$ is the covariance matrix at time $t$. $A$ is an intercept vector, $X_{jt}$ is an $S$-vector of covariate values associated with covariate $j = 1,\ldots,P$, $\beta_{j}$ is an $S$-vector of fixed effects associated with covariate $j$, and $U_{t}$ is a spatio-temporal structured effect. We use $\circ$ to denote the Hadamard product.

%The covariance matrix $\Sigma_t$ is assumed to be diagonal containing location specific %variances at each time step. We define the diagonal values as follows:
%\vspace{5mm}
%\begin{eqnarray}
%\sigma_{t, i}^{2} =  \theta_{1,i}+\theta_{2,i} Y_{t-1, i}^{2}
%\end{eqnarray}
%Note that $\sigma_{\sigma,i}^{2}$ can vary over space and time, allowing for the %potential stochastic volatility. 
%\vspace{5mm}
%\item Priors:
%
%We use informative prior distributions where possible, and conjugate distributions where %this simplifies computation as follows:
%\begin{eqnarray}
%A_{i} \sim N\left( 0, 100 \right)\\
%\beta_{1,i} \sim N\left( 0, 100 \right)\\
%\beta_{2,i \in {Dumpsite}} \sim N\left( 0, 100 \right)\\
%\beta_{2,i \in {Dredging site}} \sim N\left( 0, 0.1 \right)\\
%\beta_{3,i \in {Dumpsite}} \sim N\left( 0, 0.1 \right)\\
%\beta_{3,i \in {Dredging site}} \sim N\left( 0, 100 \right)\\
%\theta_{1,i} \sim N^{+} \left(0, 1 \right)\\
%\theta_{2,i} \sim Beta \left(1, 5 \right) \\
%\end{eqnarray}

The four different structures we consider for fitting the model involve specifying structures for the latent effects $U_t$ and the covariance matrix $\Sigma_t$. We specify prior distributions associated with these models in the section below following their definition.

\begin{description}
    \item[Model 1] An ARCH structure with varying $\Sigma_{t} = diag\{\sigma_{t, 1}^2,...,\sigma_{t, s}^2\}$ and: 
    \begin{eqnarray}
    \sigma_{t, s}^{2} &=&  \theta_{1,s}+\theta_{2,s} Y_{t-1, s}^{2} \\
    U_{t} &=& \Phi  U_{t-1}
    \end{eqnarray}
    with $\Phi = \mbox{diag}\{\phi_{1},...,\phi_{s}\}$ being an $S \times S$ diagonal matrix of autocorrelation parameters and $U_{t-1} = Y_{t-1} - A - \sum_{j=1}^{P} X_{j,t-1} \circ \beta_{j}$.
    \item[Model 2] A VAR model with a fixed time-invariant covariance matrix given an inverse-Wishart $\mathcal{IW}$ prior:
    \begin{eqnarray}
    \Sigma &\sim& \mathcal{IW}^{-1}(\Psi, \nu) \\
    U_{t} &=& \Phi  U_{t-1}
    \end{eqnarray}
    with $\nu$ and $\Psi$ as fixed hyper-parameters (we use $\nu = 14$ and $\Psi = I$ in our example), and where now $\Phi$ is a full rank matrix: 
    $$
    \Phi =\begin{bmatrix}
    \phi_{1,1}    & \hdots &  \phi_{1,s} \\
    \vdots    & \ddots  &  \vdots \\
    \phi_{s,1}  & \hdots  & \phi_{s,s}   \\
    \end{bmatrix}.
    $$
    \item[Model 3] A VARCH model which has both the full rank $\Phi$ from model 2 and the time-varying error covariance matrix of model 1.
    \item[Model 4] A VARICH integrated model that uses the difference in the latent spatio-temporal effects as well as the time-varying error covariance matrix:
    \begin{eqnarray}
    U_{t} - U_{t-1} = \Phi  (U_{t-1} - U_{t-2})
    \end{eqnarray}
    where as above the matrix $\Phi$ is of full rank.
\end{description}

To complete the model we need to specify prior distributions for all parameters. We aim to use informative priors for those where we have some degree of information, and use weakly informative and non-informative priors for the remainder. In the below we outline our prior specification for the most complex of the models we fit, model 4, though identical priors were used in the simpler models which corresponds to setting some of the parameter values to zero in a nested model structure.

Our covariates contained in $X_{j,t}$ consist of values associated with dumping (binary yes/no at each site and time), dredging (binary yes/no at each site and time), and wind speed (knots). The latter measure is common across all sites but specified separately to allow for different effects at the different sites. It is helpful, for prior specification, to consider the regression parameters $\beta$ in terms of their individual scalar components $[\beta_{dredge,s}, \beta_{dumping,s}, \beta_{wind,s}]$ at site $s$. 

Since the dumpsite and the dredging site are approximately 10 kilometers away from each other it seems reasonable to assume that dumping has no effect on turbidity at sites which are used for dredging and vice versa. Setting these values exactly to zero seemed to produce convergence problems in our Hamilton Monte Carlo Algorithm. We found that a prior of $N(0,0.3^2)$ was sufficient to remove the effects whilst still producing reasonable posterior convergence statistics. The full set of priors we used for these values is:

$$\beta_{dredge,s} \sim \left\{\begin{array}{l} N\left(0,25^2\right), \forall s: s \in \mbox{Dredging site} \\  N\left(0,0.3^2\right), \forall s: s \in \mbox{Dumpsite} \end{array}\right.$$

%- Dumping effect on turbidity at the  dumpsite:

$$\beta_{dump,s} \sim \left\{\begin{array}{l} N\left(0,25^2\right), \forall s: s \in \mbox{Dumpsite} \\ N\left(0,0.3^2\right), \forall s: s \in \mbox{Dredging site} \end{array}\right.$$

The value of 25 for the prior standard deviation of these parameters was chosen to match the approximate range of the turbidity data and was unrestricted in sign to allow for both positive and negative effects of dumping or dredging on turbidity.

For the $\Phi$ matrix we assume that off-diagonal values are most likely smaller than the diagonal ones since we believe the main autoregressive effect to be within site, and that cross site effects need to be strongly supported by the data to persist into the posterior distribution. For the diagonal terms we focus most of the prior MASS in the range (-1,1) so that the model selects for stationary behaviour, though non-stationarity can be found if the data are indicative of such phenomena. We thus use: 

$$\Phi_{ij} \sim \left\{ \begin{array}{l} N(0, 0.1^2),  \mbox{ if } i \ne j \\ N(0, 0.5^2),  \mbox{ otherwise} \end{array} \right.$$

For the remaining parameters we set:

\begin{eqnarray*}
A_{s} &\sim& N(0,100^2)\\
\beta_{wind,s} &\sim& N(0,1) \\
\theta_{1,s} &\sim& TN_{0}(0,1) \\
\theta_{2,s} &\sim& Beta(1,5)
\end{eqnarray*}

where $TN_a$ refers to the truncated normal distribution with minimum value $a$. All these are expected to be weakly informative, guiding the model towards sensible values whilst letting the data provide the majority of the information . Turbidity in our dataset ranges between 0-130 (NTU), so the prior values chosen for $A$, $\beta_{dredge}$,  $\beta_{dump}$, and $y_{missing}$ are considered to be uninformative with respect to this range. The same thing is true for $\beta_{wind}$ knowing that wind speed can reach as high as 70 knots during storms, and so a high value of 2 in units of NTU per knot (the units of $\beta_{wind}$ seems reasonable). 

As a final remark on priors we note that many of the turbidity values across sites are missing. We assume that these values are missing at random \citep[MAR;][]{little2019statistical} and impute them as part of the model fitting step by treating them as parameters to be estimated. When using the likelihood given above we found that we struggled to produce a posterior with finite variance so we added the extra prior constraint $\ y_{missing} \sim TN_{0}^{100}(0,50^2)$, a truncated normal between 0 and 100, which seemed to stabilise the missing value estimates. 

In summary, model 1 provides a baseline univariate autoregressive model with time changing variance. A more basic constant variance model was also attempted but not shown here due to poor performance. Model 2 tests whether a richer full rank vector mean structure improves the fit at the expense of the changing variance. Model 3 combines both the full vector autoregression with the time changing variance. Finally, model 4 introduces a difference in the latent parameters to capture any potential non-stationarity in the mean. Below we fit each of these models to the data described in Section \ref{data}, and use a combination of posterior predictive distributions, information criteria, and plots of the posterior distributions of the parameters to determine the optimal models which we use for interpreting our findings. 

\section{Results}\label{sec4}
In this section, we report the results of fitting the models described in Section \ref{modelling} to the turbidity data. We summarise the estimated effects of dredging and dumping operations (recall these are binary variables) and account for the wind speed effect by including the daily wind speed measured in knots. We compare the different models according to their fit to the data, and interpret the best fitting model with a view to obtaining a better understanding of turbidity behaviour in Dublin bay. 

\subsection{Model fitting and comparison}

We fit the models using R \citep{Rsoft} and the Stan modelling framework \citep{Stan}. This approach uses Hamiltonian Monte Carlo to update all parameters simultaneously and aims to rapidly converge to the posterior distribution. Through repeated fitting of the models we found that using 1000 iterations, with a warm-up period of 200 iterations, produced acceptable results. We checked convergence using the R-hat diagnostic \citep{rhat1, rhat2} which were all around the target value of 1 at convergence. 

To compare between the models, we use a combination of visual checks where we plot the one-step ahead forecasts from the model (defined via $M_t$) against the true data values, the posterior predictive distribution from the model, and more formal methods. In particular we use the Widely Applicable Information Criterion (WAIC, \cite{watanabe2012widely}) and the Leave-One-Out Information Criterion (LOO-IC, \cite{vehtari2017practical}) which penalise the likelihood of the model fit based on the complexity of the model. These two information criteria have the added advantage of being easily implemented in R and providing an uncertainty estimate on the value itself.

\begin{figure}[htbp]
\centerline{\includegraphics[scale=.4]{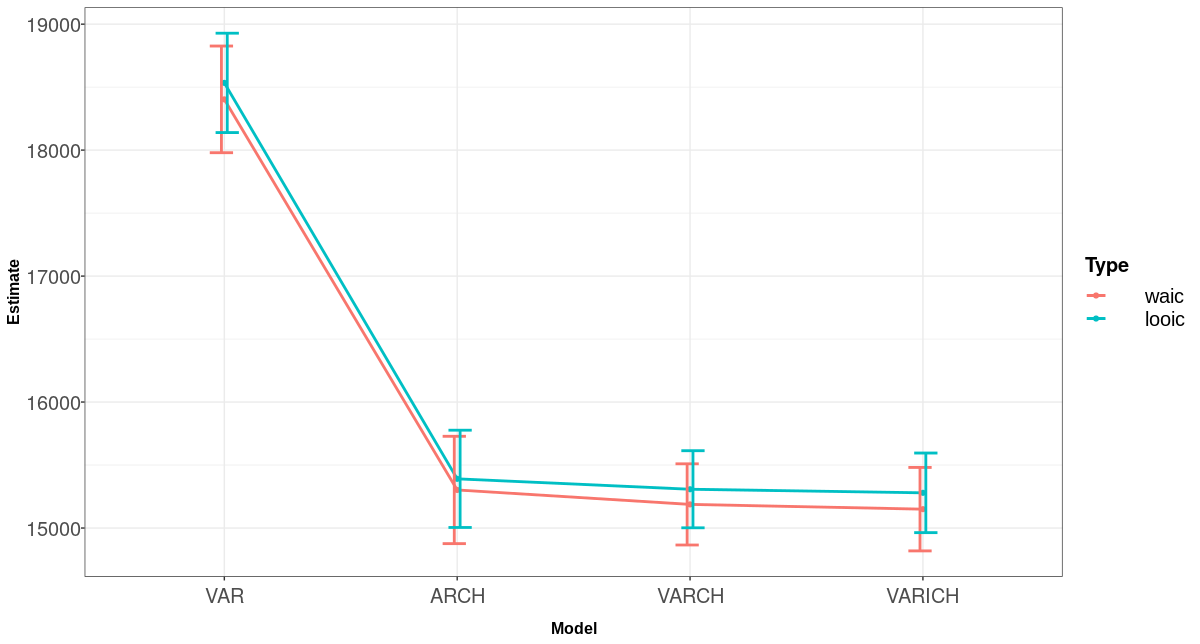}}
\caption{WAIC and LOOIC values for the four fitted models with their associated standard errors.}
\label{waic}
\end{figure}

Figure \ref{waic} shows the estimated WAIC and LOOIC values for the four models. The VARICH and VARCH models have the lowest WAIC and LOOIC values indicating better fits. However whilst the mean values of WAIC for the VARICH model are slightly lower there is no clear difference between them. The VARICH model has no extra complexity compared to VARCH, i.e. there are no extra parameters to estimate. Furthermore we computed the spectral radius of the posterior mean of $\Phi$ for both models; VARICH gave 0.33 compared to 0.98 for VARCH, which indicates that the VARICH model seems to have removed some of the non-stationarity present in the VARCH formulation. We thus use the VARICH model to create our further results. 

\begin{figure}[htbp]
\centerline{\includegraphics[scale=.45]{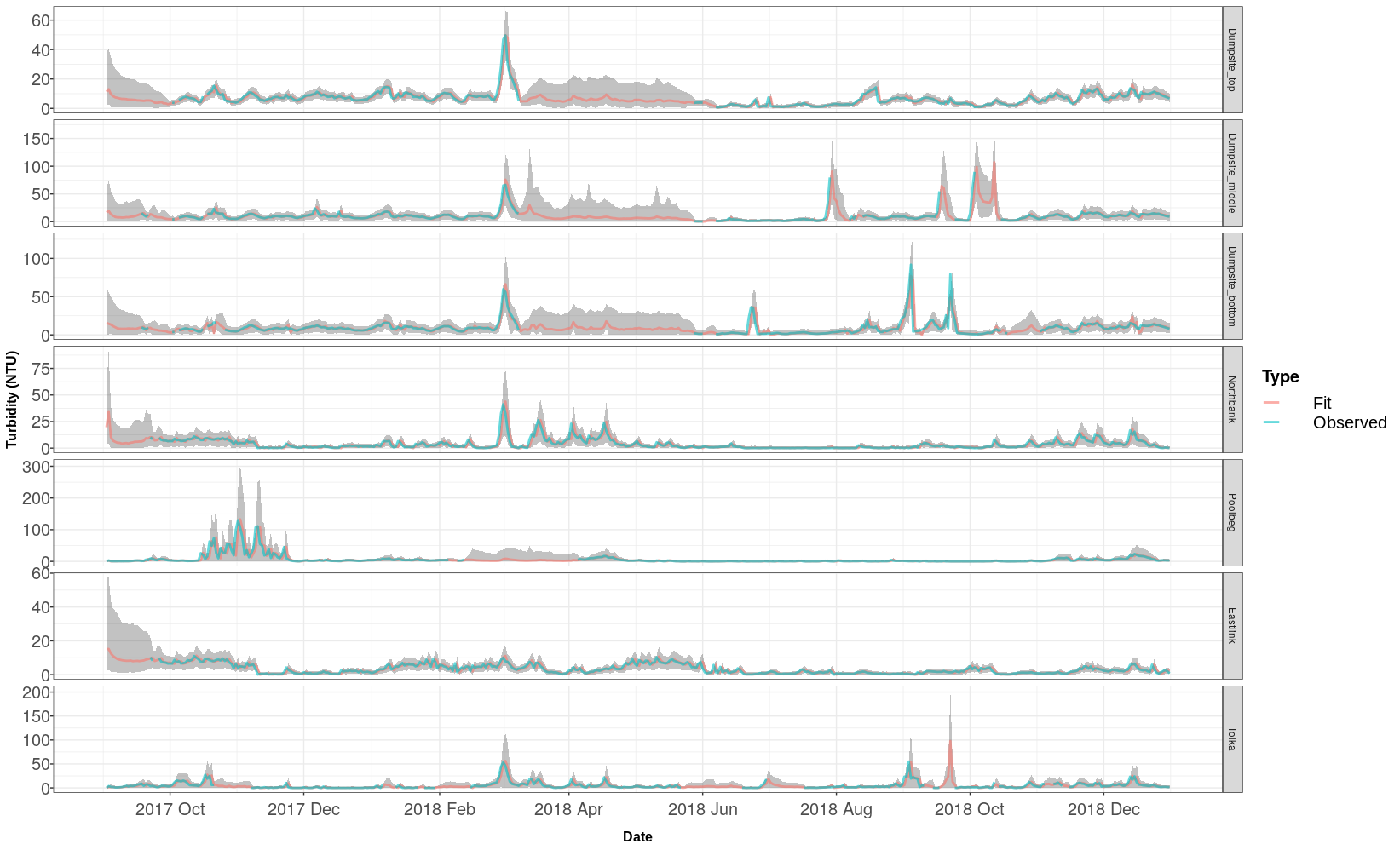}}
\caption{One step ahead forecasts from the VARICH model vs observed values of turbidity over time for the 7 buoys as labelled. Note the differing vertical axis heights. The shaded periods indicate 95\% credible intervals. }
\label{fit_obs_time}
\end{figure}

Figure \ref{fit_obs_time} shows the one step ahead fitted forecasts of turbidity from the VARICH model against the true data values. The expected values of the fit and the observed values are shown with solid lines coloured in red and blue respectively and the 95\% credible intervals are shown with grey bands. As mentioned in Section \ref{data}, the dataset has missing periods which are imputed for each location by the model during the fitting process. The vector autoregressive part of the VARICH model allows for drawing information for each site using the available information from the other sites which specifically helps regulate the uncertainty for the missing periods. As expected, the uncertainty during high volatility periods grows as expected through the dynamic structure applied to the variance.

\begin{figure}[htbp]
\centerline{\includegraphics[scale=.4]{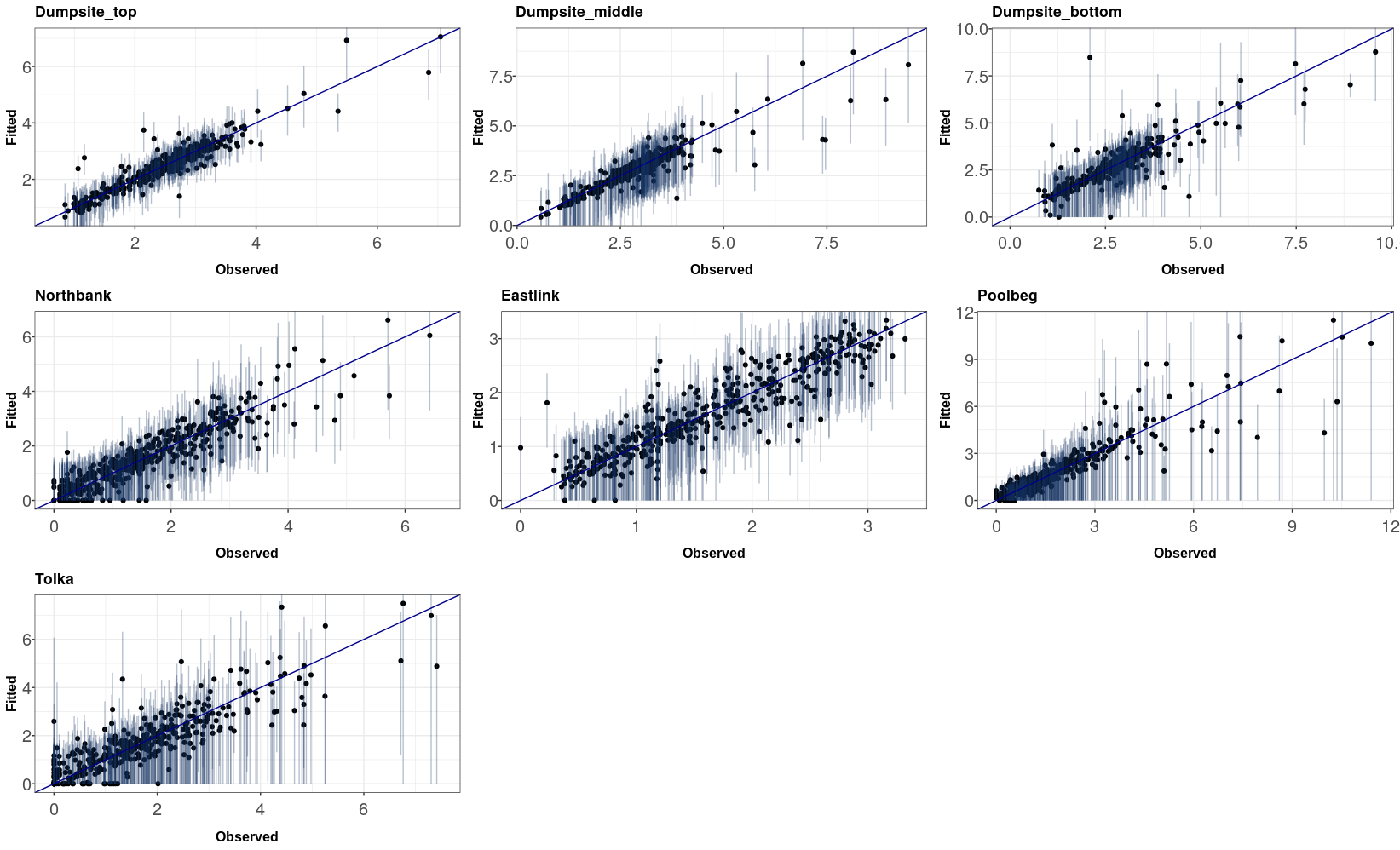}}
\caption{Fitted values from the VARICH model versus observed values of turbidity at different sites. The vertical bars indicate the 95\% uncertainty intervals which provide evidence of the coverage properties of the model.  } %Here XXX
\label{dum_fo}
\end{figure}

Figure \ref{dum_fo} shows the posterior predictive distributions from the VARICH model against the true values with vertical lines indicating the 95\% uncertainty intervals. On average the posterior prediction intervals cover 94.6\% of the data. The figure shows that the model can successfully retrieve the true values of the turbidity in the dataset with well-calibrated uncertainty estimation at the dumpsite and the dredging sites respectively.

\subsection{Effects of covariates on turbidity}

\begin{figure}[htbp]
\centerline{\includegraphics[scale=.4]{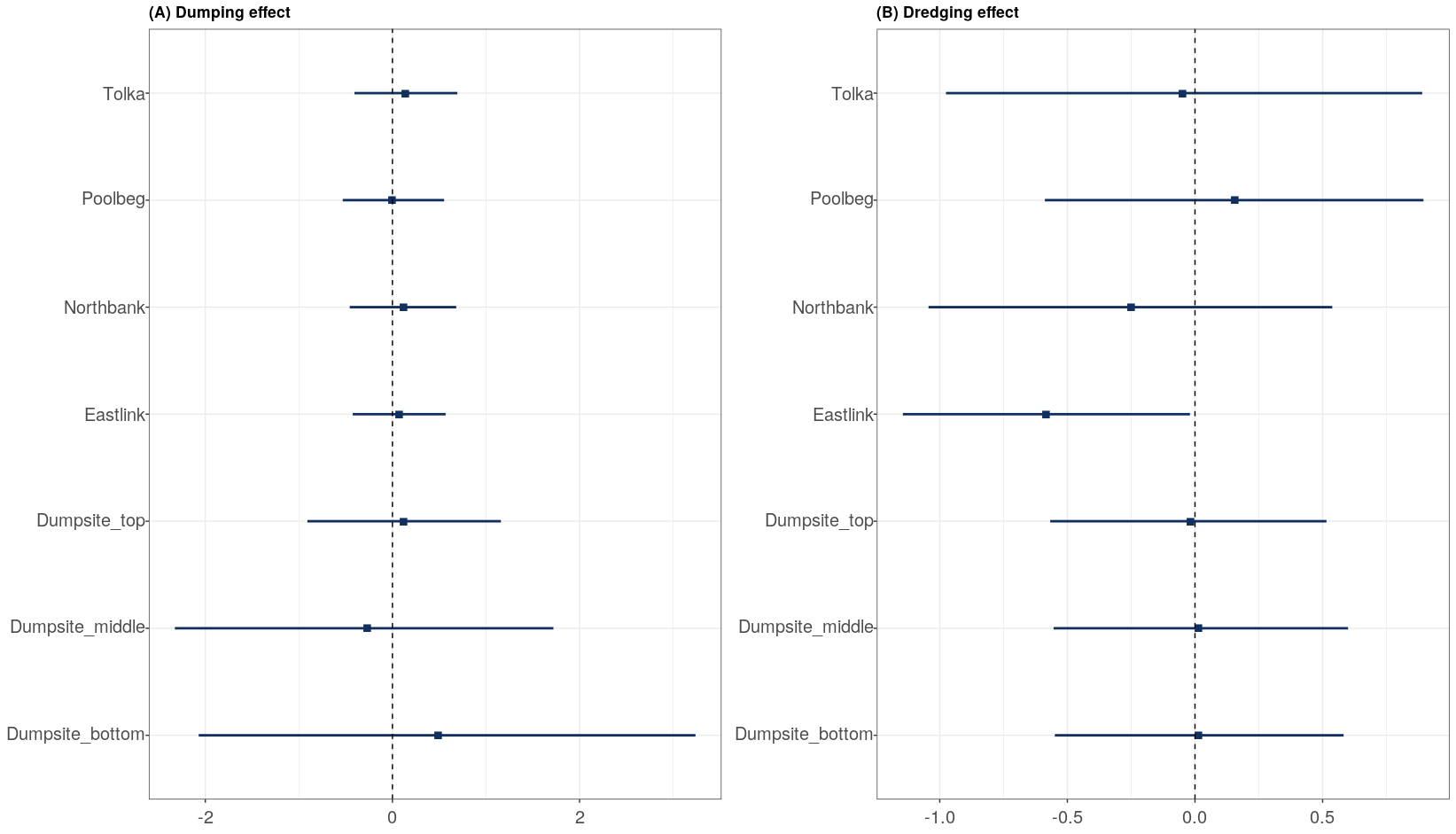}}
\caption{Dumping effect (NTU/day) at different locataions with the 95\% credible interval (A) and dredging effect (NTU/day) at different locations with the 95\% credible interval (B). }
\label{dum_ef}
\end{figure}

To determine the degree to which dumping and dredging operations affect turbidity, we evaluate the posterior distribution of the fixed effects $\beta$. Figure \ref{dum_ef} shows the expected value of the dumping and dredging effects respectively with their 95\% credible intervals for different locations. All the effects are seen to be close to zero. According to the figure, the dredging at the Eastlink site has the greatest absolute value. However, the effect is still small, and can be explained by the reduced vessel activities during dredging operations as it is the most inland of the buoys we analyse (cf Figure \ref{fig1}). An alternative explanation for this site is that freshwater flow from the Liffey and Dodder rivers can vary with the catchment zone precipitation; this impacts the salinity gradient in the water column and so potentially exerts influence over turbidity \citep{lu2020constraints}. 

\begin{figure}[htbp]
\centerline{\includegraphics[scale=.4]{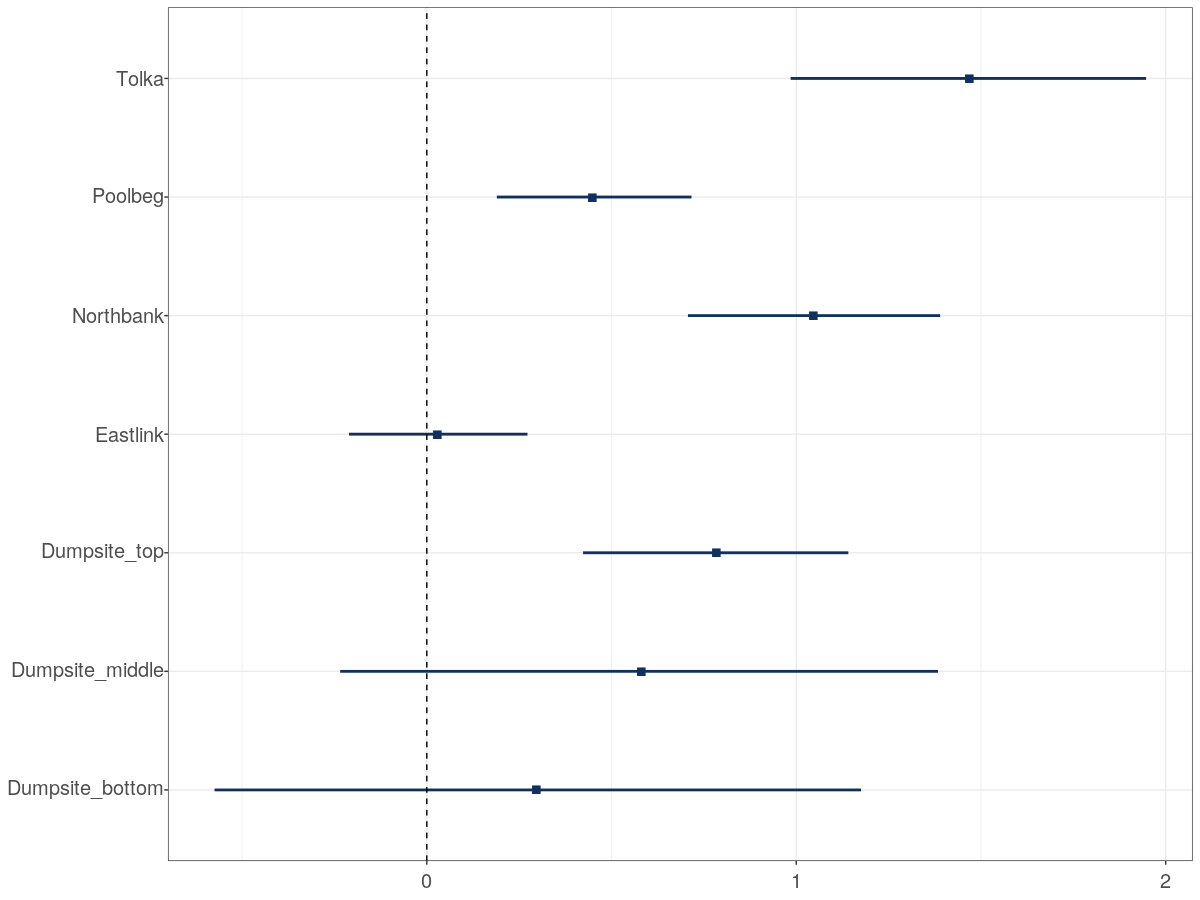}}
\caption{Effect of average wind (NTU) at different locations and depths with the 95\% credible interval.}
\label{wind_ef}
\end{figure}

Figure \ref{wind_ef} shows the wind speed effects for the 7 buoys. These are measured in NTU per knot and these wind effects can be more clearly identified than the effects of dredging and dumping. The values are reasonably consistent but with greater uncertainty at the lower positions in the dumping buoy, and a far smaller effect at Eastlink, again likely due to its position in the bay. By contrast, the Tolka buoy seems most influenced by wind and is the site that is most far out to sea. The Tolka buoy is situated  within the confines of the estuary walls,  adjacent to North Bull wall. This area of the estuary is relatively shallow and at low tide is exposed to the wind.

\subsection{Influence of the autoregressive component}

\begin{figure}[htbp]
\centerline{\includegraphics[scale=.45]{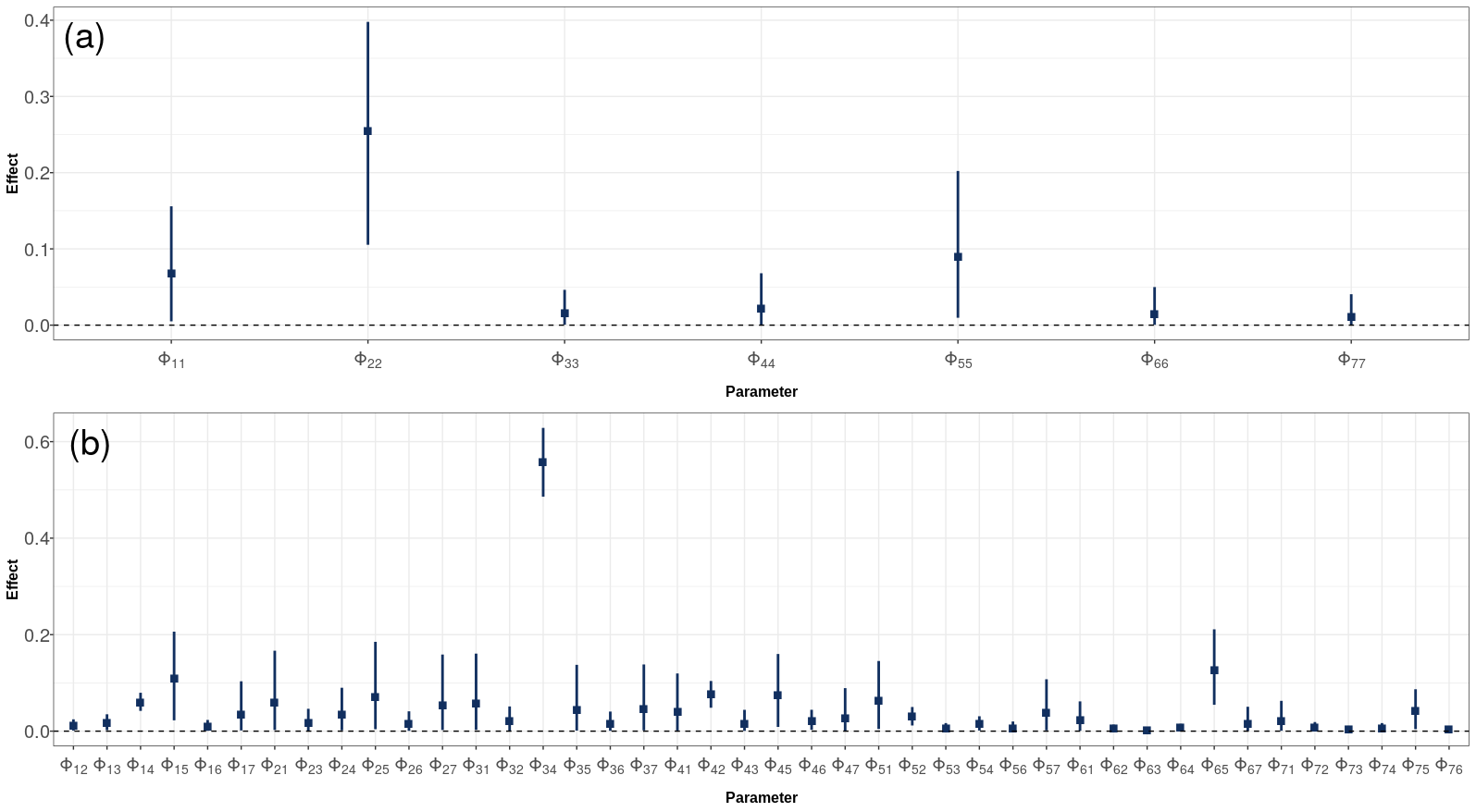}}
\caption{Coefficients of the $\Phi$ matrix with their 95\% credible interval. Diagonal values are shown in the top panel (a) and off-diagonal values are shown in (b). The two subscripts indicate the parent and child relationship respectively, so that $\Phi_{12}$ for example is the degree to which buoy 2 influences the time series of buoy 1. The numbers of the buoys follow the labelling defined in Figure \ref{fig1}.}
\label{phi_ef}
\end{figure}

As a final part of the analysis, we examine the autoregressive coefficients from the VARICH model. Figure \ref{phi_ef} shows the posterior coefficients of $\Phi$ where we have separated out the diagonal values which indicate the influence of the time series on itself from the off-diagonal elements which show the influence of one site on another. The numbering of the sites is as shown in Figure \ref{fig1}. 

Of the diagonal elements, the dumpsite (middle) seems to have the most dependence after accounting for the integration component. The other sites have values close to zero after accounting for uncertainty. Of the off-diagonal elements, some of these are well away from zero and provide for interesting, if not entirely straightforward, interpretation. $\Phi_{34}$ is the largest, corresponding to the relationship between dumpsite (bottom) and buoy 4 (Tolka), which should perhaps be read in conjunction with their joint time series behaviour as shown in Figure \ref{fit_obs_time}. Many of the other off-diagonal elements show similar clear non-zero effect sizes though they are considerably smaller than $\Phi_{34}$. These values provide evidence of cross site learning in the time series model.

\section{Discussion and conclusions}\label{disc}

We have introduced a new set of models for understanding the behaviour of turbidity in Dublin bay. The VARCH and VARICH models proposed in Section \ref{modelling} allow for measuring the effects of multiple time series on each other, whilst taking account of the known volatility changes in the time series. The combination of Bayesian modelling, VAR and ARCH structures makes our VARICH model a useful tool for flexible modelling of a wide range of real world random processes in which spatial and temporal aspects are playing major roles. Furthermore, the Bayesian approach allows for uncertainty quantification of both the fixed effects and the posterior predictions of the time series, whilst simultaneously imputing the missing values within the series. 

Our main finding has been that the dumping and dredging operations have minimal effect on the turbidity levels, which seem to be more affected by wind speed and previous values of the series. We thus suggest that, at an aggregate daily level, there is minimal effect of dredging on the turbidity levels in Dublin bay. The models we produced seem to fit the data well and the results make physical sense according to the location of the buoys in the bay. A longer time series and a more complete record would add further weight to our conclusions.

The main challenge in fitting our model has been the computational cost of the Hamiltonian MCMC. As the number of time series increases, the well known ``big n" problem of the VAR models kicks in. We have used informative priors to cope with this problem for our grouped time series, for which we have reasonable knowledge about the site connections to induce some level of sparsity in the $\Phi$ matrix structure through priors with smaller variance for the off-diagonal/cross-site effects. This is similar to \cite{santos2022bayesian}, who introduced zeros into the $\Phi$ matrix of their model for the parameters where any dependence was implausible. In the future, we may also follow \cite{yan2021} who suggested a regularization prior to make the $\Phi$ matrix sparse for faster computation.  This approach might be useful for superior estimation of the autoregression coefficients for large VAR models. A related alternative would be the recent work of \cite{heaps2022enforcing} who provide stability inducing priors to ensure that the VAR coefficients remain in a stationary region of parameter space.

\section{Acknowledgements}

We would like to thank Dr Sarah Heaps for her very help comments and assistance with the coding of some of the models. This work was supported by an SFI Investigator award (16/IA/4520). In addition, Andrew Parnell’s work was supported by: a Science Foundation Ireland Career Development Award (17/CDA/4695); a Marine Research Programme funded by the Irish Government, co-financed by the European Regional Development Fund (Grant-Aid Agreement No. PBA/CC/18/01); European Union’s Horizon 2020 research and innovation programme InnoVar under grant agreement No 818144; SFI Centre for Research Training in Foundations of Data Science 18/CRT/6049, and SFI Research Centre award 12/RC/2289\_P2. For the purpose of Open Access, the authors have applied a CC BY public copyright licence to any Author Accepted Manuscript version arising from this submission.

%big n problem: SPATIAL AND SPATIO-TEMPORAL BAYESIAN MODELS WITH R-INLA

% \section{Conclusion}\label{sec5}

% In this paper, we introduced novel approaches namely VARCH and VARICH to model the data generating processes of multivariate time series. We used a Bayesian framework for flexible model building and for conducting inference about the effects of dredging and dumping operations and wind speed on daily turbidity values at different locations and different water depths in Dublin bay. We showed that our models fit reasonably well to the time series of turbidity and are able to impute the missing values and provide us with the 95\% credible intervals for the posterior predictions. We showed that our introduced models are capable of causal inference involving multiple time series having frequent treatments introduced at different times. We compared our newly developed models to the well known models ARCH and VAR which are frequently used to analyse time series data and we showed that our models have better performance when working with dependent time series that are volatile and non stationary.

\bibliography{wileyNJD-APA.bib}
%Fix Stan reference (R should not be lowercase) 
\end{document}